\documentclass[aps, prd, twocolumn,print, showpacs, nofootinbib]{revtex4}
\usepackage{txfonts}
\usepackage{amssymb}
\usepackage{mathrsfs}
\usepackage{graphicx}
\usepackage{epsfig}
\usepackage{dcolumn}
\usepackage{bm}

\begin{document}

\title{Constraints on millicharged particles by neutron stars}

\author{Xi Huang$^{1,2,3}$, Xiao-Ping Zheng$^{1,}$\footnote{zhxp@phy.ccnu.edu.cn (Xiao-Ping Zheng), huangxi@impcas.ac.cn (Xi Huang)}, Wei-Hua Wang$^1$, and Shao-Ze Li$^1$}

\address{$^{1}$Institute of Astrophysics, Central China Normal University, Wuhan 430079, China\\
$^{2}$School of Electronic and Electrical Engineering, Wuhan Textile
University, Wuhan 430073, China\\
$^{3}$Key Laboratory of Quark and Lepton Physics (Ministry of
Education), Central China Normal University, Wuhan 430079, China\\}



\begin{abstract}
We have constrained the charge-mass ($\varepsilon-m$) phase space of
millicharged particles through the simulation of the rotational
evolution of neutron stars, where an extra slow-down effect due to
the accretions of millicharged dark matter particles is considered.
For a canonical neutron star of $M=1.4~M_{\odot}$ and $R=10~{\rm
km}$ with typical magnetic field strength $B_{0}=10^{12}$ G, we have
shown an upper limit of millicharged particles, which is compatible
with recently experimental and observational bounds. Meanwhile, we
have also explored the influences on the $\varepsilon-m$ phase space
of millicharged particles for different magnetic fields $B_{0}$ and
dark matter density $\rho_{\rm{DM}}$ in the vicinity of the neutron
star.
\end{abstract}

\maketitle

\section{INTRODUCTION} \label{S:intro}
Since Zwicky (1933) proposed the problem of the ``missing mass", the
theoretical and experimental studies on dark matter (DM) have
attracted more and more attention. It is well-known that the total
energy of the Universe contains about 27\% DM and about 5\% baryon
matter as well as about 68\% dark energy according to the most
recent cosmological results based on Planck measurements of the
cosmic microwave background (CMB) temperature and lensing-potential
power spectra \citep{Ade14}. In recent years, experimental
physicists have concentrated their attention on the direct and
indirect detections of DM particles. Direct search experiments aim
to detect individual interactions, i.e., DM particles scatter off
target nuclei of detectors, such as DAMA \citep{Savage09}, CDMS II
\citep{Ahmed10}, CoGeNT \citep{Aalseth11}, CRESST-II
\citep{Angloher12}, XENON100 \citep{Aprile12}, LUX \citep{Akerib14},
PandaX \citep{Xiao14} and so on. Indirect detection experiments
mainly probe annihilation or decay products (gamma-rays, neutrinos
and charged cosmic rays) of DM, such as AMS-02, PAMELA, Fermi-LAT,
IACTs, IceCube, ANTARES, Super-K \citep{Rott13,Conrad14}. From a
theoretical point of view, several candidates of DM particles are
supposed, for example, neutralinos \citep{Jungman96,Bertone05},
Majorana neutrinos and technibaryons \citep{Gudnason06,Kouvaris07}.
In this paper, we are interested in another candidate of DM in the
form of millicharged (MC) particles
\citep{Goldberg86,Kouvaris13,Kouvaris14}.

The electric charge of all the particles in the Standard Model
appears to be an integer multiple of $e/3$, where $e$ is the charge
of the electron. However, MC particles have electric charge
$e'=\varepsilon e$, where $\varepsilon$ is any real number and
$\varepsilon <1$. They can be either bosons or fermions. MC
particles were first proposed in order to solve the DM puzzle a long
time ago \citep{Goldberg86,De Rujula90,Dimopoulos90}. Hereafter
millicharged dark matter particles (MCDM) have been studied
extensively (see Ref. \citep{Kouvaris13} and references therein). It
is worth noticing that Huh {\it et al.} \citep{Huh08} presented a
possible explanation of the 511 KeV galactic $\gamma$-ray due to
positron productions from MCDM. MC particles have an obvious impact
on the standard picture of the Universe in many ways. Firstly MC
particles can significantly influence the expansion rate of the
Universe and the baryon-to-photon ratio during the epoch of big bang
nucleosynthesis (BBN) in the early Universe
\citep{Davidson00,Melchiorri07,Berezhiani13a}. Secondly, MC
particles also may explain the creation of galactic magnetic fields
at the cosmological epoch of the galaxy formation
\citep{Berezhiani13b}. Thirdly, the anisotropy power spectrum of the
CMB is affected by MC particles in several respects
\citep{Dubovsky01,Dolgov13}.

At present MC particles are already constrained by experimental and
observational data
\citep{Prinz98,Davidson00,Langacker11,Jaeckel13,Berezhiani13a},
including the $\varepsilon-m$ ($m$ is the mass of MC particle) phase
space and the fraction of MC particles. These existing constraints
on MC particles mainly contain laboratory bounds (including the
limits from experiments searching for MC particles
\citep{Mitsui93,Prinz98}), cosmological bounds (BBN, CMB anisotropy)
and constraints from stellar evolution (globular clusters, white
dwarfs, red giants, and supernova 1987A). For very light MC
particles, $m<$ 1 keV, the reactor experiments show a strict limit
$\varepsilon<10^{-5} $ \citep{Gninenko07}. For MC particles that are
lighter than electrons, $m<m_e$, the best particle physics bound
$\varepsilon<3.4\times10^{-5}$ is given from the data that
ortho-positronium decay to dark matter pairs \citep{Badertscher07}.
For heavier MC particles, $m>m_e$, the bounds become weaker
\citep{Prinz98}. For $m=$ 1 MeV, the bound is
$\varepsilon<4.1\times10^{-5}$, and for $m=$ 100 MeV, the bound is
up to $\varepsilon<5.8\times10^{-4}$. If the MC particles become
heavier than 100 MeV, the bound $\varepsilon\sim10^{-2}$ is allowed,
while for $m>$ 1 GeV it can be as large as $\varepsilon=0.1$.

With the discovery of neutron star (NS) by Bell and Hewish
\citep{Hewish68}, the properties of NS have been studied by many
theoretical and experimental physicists
\citep{Glendenning97,Page06,Lattimer07,Potekhin10}. NS is a compact
star, which is characterized by strong gravitational field,
electromagnetic field, extreme strong and weak interaction. Thus it
is usually regarded as a ``natural laboratory'' with extreme
physical condition. At present it seems that all currently observed
pulsar periods mainly lie between milliseconds and a few seconds. As
it can be seen from the $P-\dot{P}$ ($P$ and $\dot{P}$ are the spin
period and its time derivative, respectively) diagram (see Fig. 1),
most radio pulsars have a period of around one second and the period
derivatives in the range of $10^{-16}$ to $10^{-14}$ s s$^{-1}$.
However, one group of sources having higher magnetic field
($10^{13-15}$ G), relatively long periods ($>$1 s), and high period
derivatives ($10^{-13}-10^{-9}$ s s$^{-1}$ ) are magnetars. The
other group of sources are millisecond pulsars with old age
($\sim10^{9}$ yr) and lower magnetic field ($10^{8-9}$ G). In the
ATNF pulsar catalogue \citep{Manchester05}, the longest spin period
for normal radio pulsars to date is $\sim$8.51 s of PSR J2144-3933
whose characteristic age is $2.72\times10^{8}$ yr and surface
magnetic field is $2.08\times10^{12}$ G \citep{Young99}. Except for
short-lived glitches \citep{McCulloch83}, the rate of rotation of a
NS drops steadily as a function of time, especially for millisecond
pulsars which have very low values of $\dot{P}$ and very low timing
noise. It is generally believed that the spin-down of the NS is due
to magnetic dipole radiation (MDR) and electromagnetic torques in
the magnetosphere \citep{Contopoulos06}.

Limits on DM from neutron stars have been deeply investigated in a
large amount of literatures so far
\citep{Bertone08,Kouvaris08,Kouvaris10,de
Lavallaz10,Fan11,Kouvaris11,Kouvaris12,Huang14,Zheng14}. Recently,
Kouvaris and Perez-Garcia \citep{Kouvaris14} have discovered that
the electric charges could be expelled from the star as MCDM are
accreted onto the NS. The escaping charged particles will provide an
extra current, thus the torque is produced. They find that this
mechanism yields an extra spin-down of neutron stars and the braking
indices can be substantially smaller than 3 predicted by the MDR
model.

We follow the philosophy of Kouvaris and Perez-Garcia
\citep{Kouvaris14}, but explore how to constrain the $\varepsilon-m$
of MC particles via the currently observed pulsar periods. We
consider the effect of the additional torque by the extra currents
due to the accretions of MCDM onto the NS on the period of pulsars.
We also impose a limit of pulsar's period in the $P-\dot{P}$ diagram
on the period's evolution equation as a cut-off. By doing so, the
$\varepsilon-m$ of MC particles will be constrained under the given
DM density.

This paper is organized as follows. Firstly, we revisit the
mechanism of the accretions of MCDM onto the NS (electromagnetic
accretion), then the rotational evolution equations of a NS
considered the MDR and the torques, which are produced by the
Goldreich and Julian (GJ) current \citep{Goldreich69} $I_{\rm GJ}$
and the extra current $I_{\rm DM}$, are shown in Section 2.
Secondly, we numerically show the bounds of the $\varepsilon-m$ of
MC particles in Section 3. Finally, we present the conclusions in
Section 4.

\section{THE MODELS}
First we show how the accretions of MCDM onto the NS can lead to an
excess of currents \citep{Kouvaris14}. If the Larmor radius $r_{\rm
L}=\frac{m\upsilon_\perp}{\varepsilon eB}$ of MCDM is much smaller
than the curvature radius $R_{\rm curv}=|\frac{1}{B}\frac{{\rm
d}B}{{\rm d}r}|^{-1}$, it may follow the magnetic field lines in the
magnetosphere within the light cylinder of the NS and cross the
surface of the star. $\upsilon_\perp$ is the particle's velocity
perpendicular to the magnetic field $B$. As the accumulations of
MCDM trapped in the NS, once the Coulomb repulsive force becomes
larger than the gravitational force, the electric charges will be
expelled from the star. The expelled charges are in the form of
either electrons or protons, which depend on the charge of the
accreted MCDM. Once the equilibrium between the accretions of MCDM
and the expulsions of electric charges has been established, stable
extra currents will be formed.

As stated previously, the slow-down of the rotation powered pulsars
is due to the MDR and the braking torques provided by the outflowing
plasmas \citep{Kramer06}. The rotational kinetic energy loss rate of
the NS is $\dot E=\mathcal{I}\Omega\dot \Omega=-L$, where
$\mathcal{I}$ and $\Omega$ are the moment of inertia and the angular
velocity of the star, respectively, and $L$ can be expressed as two
components \citep{Harding99}
\begin{equation}
L=L_{\rm orth}\sin^{2}\theta+L_{\rm align}\cos^2\theta.
\end{equation}
In the above equation, $L_{\rm orth}$ represents the ``orthogonal"
case that the angular momentum of the star is decreased by the MDR,
$L_{\rm align}$ represents the ``aligned" case that the torque is
produced by the electric current from escaping charged particles
that follow the open magnetic-field lines in the magnetosphere of
the NS, and $\theta$ is the angle between the rotational and the
magnetic axes (magnetic inclination angle). The two components can
be written as
\begin{equation}
L_{\rm orth}=\frac{B_{0}^{2}\Omega^{4}R^6}{4c^3},\hspace{1cm} L_{\rm
align}=\frac{B_{0}\Omega(\Omega-\Omega_{\rm death})R^{3}I}{2c^2},
\end{equation}
where $B_{0}$ and $R$  are the magnetic field strength on the polar
cap (the open magnetic-field lines in the light cylinder connect to
the surface of the star) and radius of the star, respectively, $c$
is the light speed, and $\Omega_{\rm death}$ is the angular velocity
below which the pulsar emission dies. The total current $I$ can be
written as the sum of $I_{\rm GJ}$ and $I_{\rm DM}$, $I=I_{\rm
GJ}+I_{\rm DM}$, where $I_{\rm GJ}$ represents the emission of
relativistic charged particles from the surface cap regions ($\sim
\pi R_C^2$, $R_C$ is the polar cap radius) of the NS, due to a
particle acceleration gap potential $V_{\rm{gap}}$ of the order of
$10^{12}$ V \citep{Hibschman01} that develops along open magnetic
field lines in the vicinity of the polar cap, and $I_{\rm DM}$
represents the extra current from the expelled electric charges
(electrons or protons) due to the accretions of MCDM onto the NS.

According to the pioneering work of Goldreich and Julian
\citep{Goldreich69}, the GJ current is $I_{\rm GJ}=\pi
R_C^2\rho_{\rm GJ}c$, where $\rho_{\rm GJ}$ is the GJ charge
density. The polar cap radius can be expressed as
$R_C=R\sin{\theta_{\rm P}}=R(R/R_{\rm L})^{1/2}$, where $\theta_{\rm
P}$ is the angle of the polar cap to the center of the star and
$R_{\rm L}=cP/2\pi$ is the light cylinder radius \citep{Sturrock71}.
In the polar cap region, the GJ charge density is given by
\citep{Goldreich69}
\begin{equation}
\rho_{\rm
GJ}=7\times10^{10}e\left(\frac{B_{0}}{10^{12}~\rm{G}}\right)\left(\frac{\rm~s}{P}\right)~\rm{cm^{-3}},
\end{equation}
thus, we can easily deduce
\begin{equation}
I_{\rm
GJ}\simeq1.4\times10^{30}e\left(\frac{B_{0}}{10^{12}~\rm{G}}\right)\left(\frac{\rm~s}{P}\right)^2~\rm{s^{-1}}.
\end{equation}
For the extra current $I_{\rm DM}$ due to the accretions of MCDM
onto the NS, we directly adopt the results in Ref.
\citep{Kouvaris14},
\begin{equation}
I_{\rm DM}\simeq1.0\times10^{29}\varepsilon e\left(\frac{\rho_{\rm
DM}}{0.3~\rm{GeV}~\rm{cm^{-3}}}\right)\left(\frac{1~\rm{GeV}}{m}\right)\left(\frac{P}{\rm~s}\right)~\rm{s^{-1}}.
\end{equation}

\begin{figure}
\centering
\includegraphics[scale=0.33]{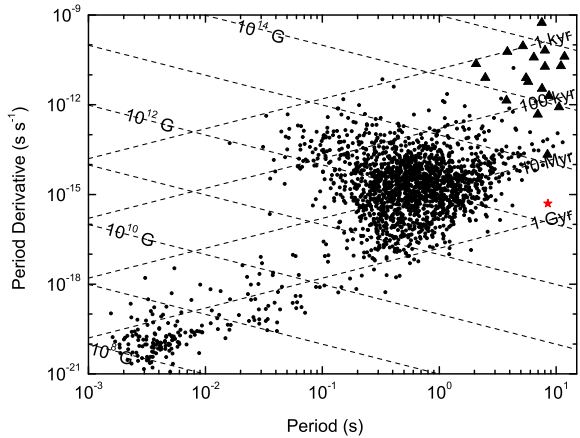}
\caption{Distribution of all known pulsars on the $P-\dot{P}$
diagram. Dots are rotation powered pulsars (including normal pulsars
and millisecond pulsars, which dominate the pulsar population) (from
ATNF: http://www.atnf.csiro.au/research/pulsar/psrcat/). Triangles
are magnetars (from McGill magnetar catalog:
http://www.physics.mcgill.ca/$\sim$pulsar/magnetar/main.html). The
red star denotes the famous PSR J2144-3933, which has the longest
spin period for normal radio pulsars. The dash lines of pulsar
characteristic age are defined by ${P}/({2\dot{P}})$ and another
dash lines of surface dipole magnetic field are conventionally
defined as $3.2\times10^{19}(P\dot{P})^{1/2}$, which are based on
the MDR.} \label{fig:1}
\end{figure}

Finally, using the Eqs. (1) and (2), we can get the rotational
evolution equation of the NS
\begin{equation}
\frac{{\rm d}\Omega}{{\rm
d}t}=-\frac{B_{0}^{2}\Omega^{3}R^{6}}{4\mathcal{I}c^3}\left[\sin^{2}\theta+\left(1-\frac{\Omega_{\rm
death}}{\Omega}\right)\left(1+\frac{I_{\rm DM}}{I_{\rm
GJ}}\right)\cos^{2}\theta \right].
\end{equation}
We define $\lambda=\frac{I_{\rm DM}}{I_{\rm GJ}}$ for convenience.
As we can see from Eq. (6), it easily recover the scene of the MDR
for $\lambda=0$ (no extra current). It is worth noticing that we can
take $\frac{\Omega_{\rm death}}{\Omega}\ll1$ safely, since normal
pulsars couldn't be older than $10^7$ to $10^8$ years. We compare
the evolution curve of period from Eq. (6) with that from the scene
of pure MDR. Fig. 2 shows the differences between them. It is
obvious from Fig. 2 that the differences become larger gradually as
the increases of age. We use the 8.51 s period at $2.72\times10^{8}$
yr (which are the observational data of PSR J2144-3933) as the
cut-off of neutron stars, because it is the longest observed period
of normal pulsars and possible longer period should fall below the
so-called ``death line" on the $P-\dot{P}$ diagram, where pulse
emission can't be observed. As shown in Fig. 1, the location of PSR
J2144-3933 on the $P-\dot{P}$ diagram indicates that the star could
be a good candidate as the cut-off for the period evolution and the
characteristic age of normal radio pulsars. However, both the
millisecond pulsars and magnetars are unsuitable to constrain the
$\varepsilon-m$ of MC particles. It is generally believed that the
millisecond pulsars have a complicated and turned evolution.
Recycling neutron stars to millisecond periods may be a key process.
Additionally, the extra current $I_{\rm DM}$ is much smaller than
the corresponding GJ current $I_{\rm GJ}$ according to Eq. (4) for
millisecond pulsars. The tiny differences have no imprint in Fig. 2
even if we overlook the recycled processes. For higher magnetic
fields, there are only 28 currently known magnetars and magnetar
candidates \citep{Olausen14}. The physics of magnetars is unclear so
far. Their emission is probably powered by the non-rotational
kinetic energy or by their decay of super-strong magnetic field
\citep{Goldreich92,Heyl98}. The pure MDR model may not be a good
description for magnetars. Therefore, we think that it is seriously
inaccurate and unreliable to constrain the parameters of MC
particles by magnetars.

\begin{figure}
\centering
\includegraphics[scale=0.33]{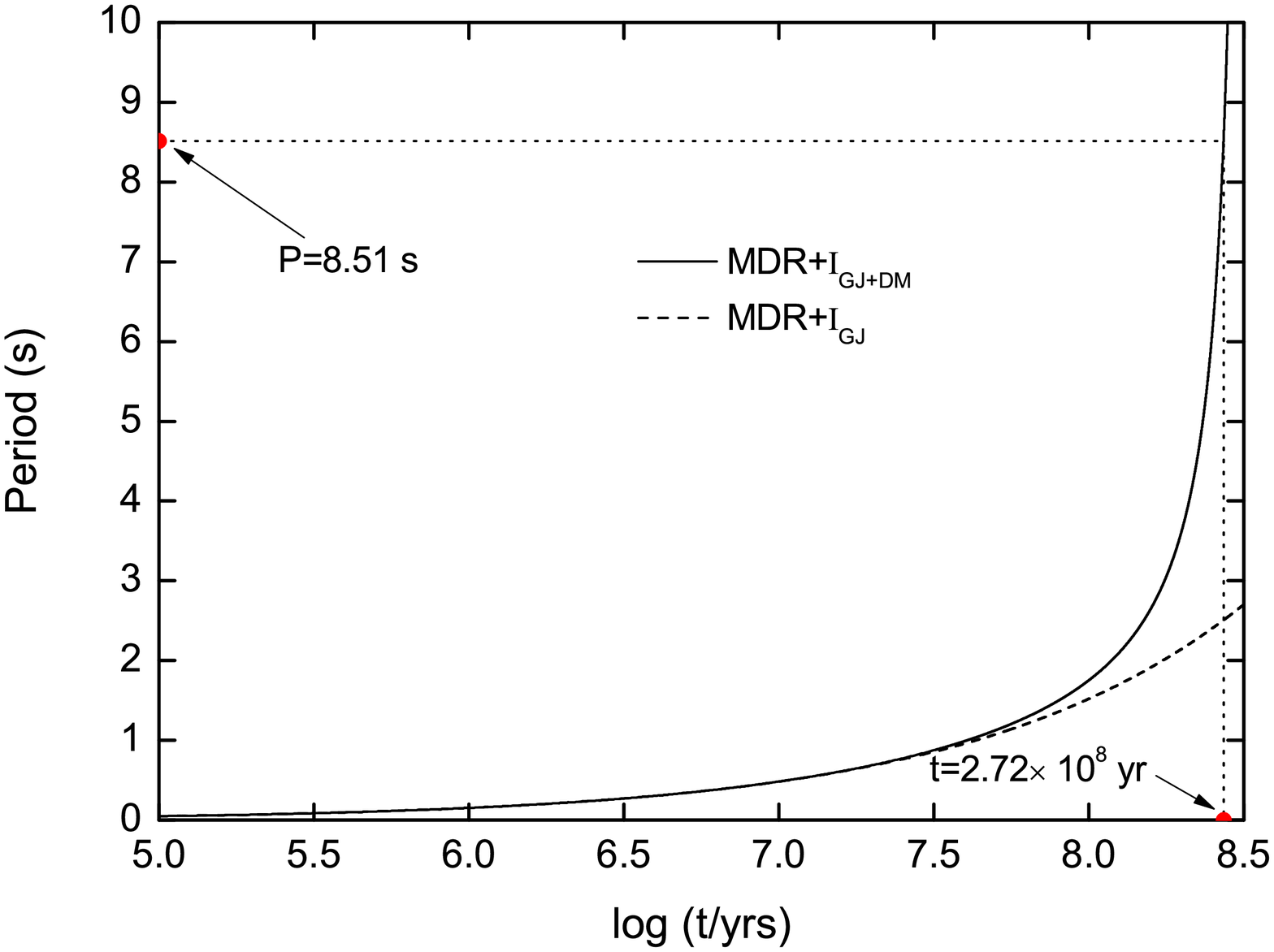} \caption{The period
evolution of a canonical NS with $M=1.4~M_{\odot}$ and $R=10~{\rm
km}$ for $B_{0}=10^{12}$ G. The dash curve represents the spin-down
induced by pure MDR ($\lambda=0$ in Eq. (6)). The solid curve shows
an extra slow-down of the NS with magnetic inclination angle
$\theta=45^{\circ}$ by the additional torque due to the accretions
of MCDM with $m=1$ GeV and $\varepsilon=5.04\times10^{-2}$. The
labels for the red dots represent the 8.51 s period at
$2.72\times10^{8}$ yr.} \label{fig:2}
\end{figure}

\section{NUMERICAL RESULTS AND DISCUSSIONS} \label{Section III}

\begin{figure*}
\centering \includegraphics[width=12cm,height=9cm]{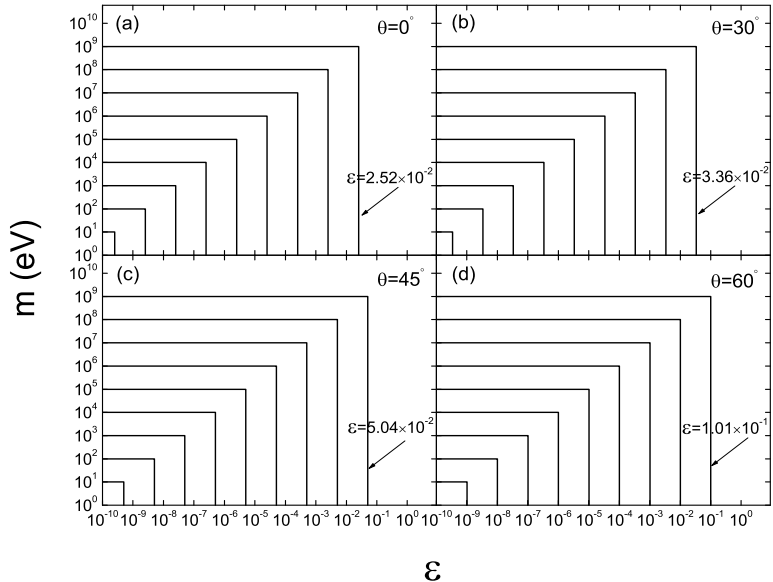}
\caption{The $\varepsilon-m$ phase space of MC particles constrained
by a canonical NS of $M=1.4~M_{\odot}$ and $R=10~{\rm km}$ with
$B_{0}=10^{12}$ G for different magnetic inclination angles.}
\label{fig:3}
\end{figure*}

For convenience, we assume that the star consists of nonsuperfluid
neutrons, protons and electrons, and suppose that the density of the
star is uniform. In our calculations, we consider a canonical NS of
$M=1.4~M_{\odot}$ and $R=10~{\rm km}$, and take the initial period
$P_0=1\hspace{0.1cm}{\rm ms}$, the moment of inertia of the star
$\mathcal{I}=\frac{2}{5}MR^{2}\approx10^{45}~\rm{g~cm^2}$, and the
local DM density in the vicinity of the NS
$\rho_{\rm{DM}}=~100\times0.3~\rm{GeV/cm^3}$ ($0.3~\rm{GeV/cm^3}$
represents the standard DM density around the Earth).

The $\varepsilon-m$ phase space of MC particles could be constrained
using the following strategy. First, we plug a great number of
groups of ($\varepsilon,m$) into Eqs. (4), (5) and (6) for various
magnetic field strength and different magnetic inclination angles
under the given DM density in the vicinity of the NS. Then, we
simulate Eq. (6) up to $2.72\times10^8$ yr as the cut-off for the
age of the NS. If the evolution of the rotational period of the star
exceeds the longest observed period of normal radio pulsars to date
($\sim$8.51 s), we consider this group of ($\varepsilon,m$) invalid.

The $\varepsilon-m$ phase space of MC particles constrained by a
canonical NS of $M=1.4~M_{\odot}$ and $R=10~{\rm km}$ with typical
$B_{0}=10^{12}$ G for different magnetic inclination angles is shown
in Fig. 3. It can be seen from Fig. 3(a) that the charge of MC
particles can reach up to $\varepsilon=2.52\times10^{-2}$ for
heavier particles with $m=$ 1 GeV, the bound
$\varepsilon>2.52\times10^{-5}$ is ruled out apparently for MC
particles with $m=$ 1 MeV, and the limit
$\varepsilon\leq2.52\times10^{-8}$ is allowed for lighter particles
with $m=$ 1 KeV. Furthermore, if the mass $m$ decreases by one order
of magnitude, the charge $\varepsilon$ would reduce by the same one
order of magnitude. As shown above, the lighter the mass, the
smaller the charge. It is obvious from Eq. (6) that the rotational
angular velocity of the star could decrease rapidly as $\lambda$
increases for the same magnetic field strength $B_{0}$, which result
in the rise of the spin period even to the unreasonable value. Fig.
3(b), (c) and (d) show that the upper limit of $\varepsilon$ is
inversely proportional to $\cos^2\theta$ for the same mass $m$ of MC
particles (see Eq. (6)). To summarize, we would present an upper
limit of MC particles,
$(\frac{1~\rm{GeV}}{m})\times\varepsilon\leq2.52\times10^{-2}/\cos^2\theta$,
which is consistent with experimental and observational bounds
\citep{Gninenko07,Badertscher07,Prinz98}.

It is worth stressing that, as shown in Eq. (6), the rate of
deceleration of the rotation $\dot{\Omega}$ also depends on the
magnetic field strength $B_{0}$. For $B_{0}=10^{11}$ G, the bounds
become weaker, the charge $\varepsilon$ of MC particles can't be
constrained for heavier particles with $m=$ 1 GeV, however up to
$\varepsilon\leq3.26\times10^{-3}/\cos^2\theta$ for $m=$ 1 MeV and
$\varepsilon\leq3.26\times10^{-6}/\cos^2\theta$ for lighter
particles with $m=$ 1 KeV. On the other hand, it can be seen from
Eq. (5) that, the extra current $I_{\rm DM}$ relies on the DM
density $\rho_{\rm{DM}}$ in the vicinity of the NS. Although
$\rho_{\rm{DM}}$ around the NS is uncertain to date, we could draw a
conclusion that, the larger the DM density, the smaller the charge
for the MC particles with the same mass, e.g., if the DM density
$\rho_{\rm{DM}}$ increases by one order of magnitude, the charge
$\varepsilon$ would decrease by one order of magnitude
correspondingly for the same mass $m$ of MC particles.

\section{CONCLUSIONS} \label{Section IV}
We have constructed the constraints of the $\varepsilon-m$ of MC
particles by neutron stars based on an enhanced slow-down of neutron
stars due to an extra current yield by the accretions of MCDM. For a
canonical NS of $M=1.4~M_{\odot}$ and $R=10~{\rm km}$ with typical
magnetic field strength $B_{0}=10^{12}$ G, we have shown an upper
limit of MC particles,
$(\frac{1~\rm{GeV}}{m})\times\varepsilon\leq2.52\times10^{-2}/\cos^2\theta$,
which indicates the charge would be smaller as the mass of MC
particles becomes lighter and is compatible with experimental and
observational bounds. The specific limits are as follows, the charge
of MC particles could rise to
$\varepsilon=2.52\times10^{-2}/\cos^2\theta$ for heavier particles
with $m=$ 1 GeV, the bounds
$\varepsilon\leq2.52\times10^{-5}/\cos^2\theta$ and
$\varepsilon\leq2.52\times10^{-8}/\cos^2\theta$ are allowed for the
MC particles with $m=$ 1 MeV and $m=$ 1 KeV, respectively, i.e., if
the mass $m$ decreases by one order of magnitude, the charge
$\varepsilon$ would reduce by the same one order of magnitude.
However, for the NS with lower magnetic fields ($10^{10-11}$ G), the
bounds become weaker. In particular, for millisecond pulsars
($10^{8-9}$ G) and magnetars ($10^{13-15}$ G), it is not appropriate
to bound the $\varepsilon-m$ of MC particles. In addition, we have
also investigated the influence on the $\varepsilon-m$ of MC
particles for the different DM density $\rho_{\rm{DM}}$ in the
vicinity of the NS. It is obvious that, the larger the DM density,
the smaller the charge for the MC particles with the same mass.

The model we adopted is uniform stellar configuration. However, it
is well-known that the neutron stars constructed by the realistic
equations of states can be approximated as the uniform case. Our
work have shown an upper limit of the $\varepsilon-m$ of MC
particles based on the new mechanism, and the results will be
unchanged in the order of magnitude when considering the realistic
equations of states.

\section*{\uppercase {acknowledgments}}
The authors would like to thank the referee very much for helpful
comments and Yunwei Yu for useful discussions, which have
significantly improved our work. We also thank Shuhua Yang for help
in English writings. This work is supported by the National Natural
Science Foundation of China (Grant No. 11178001) and CCNU-QLPL
Innovation Fund (Grant No. QLPL2015P01).

\end{document}